\documentclass[a4paper,11pt]{article}
\usepackage{pos}
\usepackage{tikz}
\usetikzlibrary{shapes,arrows,positioning}

\title{Camera Calibration for the IceCube Upgrade and Gen2}
\ShortTitle{Camera Calibration for the IceCube Upgrade and Gen2}

\author{The IceCube Collaboration \\{\normalsize \normalfont(a complete list of authors can be found at the end of the proceedings)}}

\emailAdd{woosik.kang@icecube.wisc.edu}
\emailAdd{ctoennis@icecube.wisc.edu}

\abstract{An upgrade to the IceCube Neutrino Telescope is currently under construction. For this IceCube Upgrade, seven new strings will be deployed in the central region of the 86 string IceCube detector to enhance the capability to detect neutrinos in the GeV range. One of the main science objectives of the IceCube Upgrade is an improved calibration of the IceCube detector to reduce systematic uncertainties related to the optical properties of the ice. We have developed a novel optical camera and illumination system that will be part of 700 newly developed optical modules to be deployed with the IceCube Upgrade. A combination of transmission and reflection photographic measurements will be used to measure the optical properties of bulk ice between strings and refrozen ice in the drill hole, to determine module positions, and to survey the local ice environments surrounding the sensor module. In this contribution we present the production design, acceptance testing, and plan for post-deployment calibration measurements with the camera system.

\vspace{4mm}
{\bfseries Corresponding authors:}
Woosik Kang$^{1*}$, Jiwoong Lee$^{1}$, Gerrit Roellinghoff$^{1}$, Carsten Rott$^{1,2}$, Christoph T\"onnis$^{1,3}$\\
{$^{1}$ \itshape Department of Physics, Sungkyunkwan University, Suwon 16419, Korea}\\
{$^{2}$ \itshape Department of Physics and Astronomy, University of Utah, Salt Lake City, UT 84112, USA}\\
{$^{3}$ \itshape Institute of Basic Science, Sungkyunkwan University, Suwon 16419, Korea}\\[4mm]

$^*$ Presenter

\FullConference{37$^{\rm{th}}$ International Cosmic Ray Conference (ICRC 2021)\\
		July 12th -- 23rd, 2021\\
		Online -- Berlin, Germany}

}

\begin{document}

\maketitle

\section{Introduction}
\label{introduction}
Located at the geographic South Pole, the IceCube Neutrino Observatory~\cite{ICdetector} is the world's largest neutrino telescope in terms of instrumented volume. It consists of a Cherenkov detector of one cubic kilometre volume using the ultra-pure Antarctic ice~\cite{SPICE} at depths between 1.45~km and 2.45~km and a square kilometre air-shower detector at the surface of the ice~\cite{IceTop}. The primary objectives of the detector are the measurement of high-energy astrophysical neutrino fluxes and determining the sources of these fluxes~\cite{HEnu_Science,TXS}.\\
\indent Currently an upgrade comprised of seven densely instrumented strings in the centre of the active volume of the IceCube detector with new digital optical modules (DOMs) is being built as the IceCube Upgrade~\cite{ICRC2019:ICU-project}. On each string DOMs will be regularly spaced with a vertical separation of 3\,m between depths of 2160\,m and 2430\,m below the surface of the ice. Three different types of DOMs will be used: The pDOM which is based on the design of the existing IceCube DOMs with upgraded electronics, the D-Egg~\cite{ICRC2021:D-Egg} which has two 8-inch photomultiplier tubes (PMTs) (facing up and down respectively), and the mDOM~\cite{ICRC2021:mDOM} which has 24~three-inch PMTs distributed for close to uniform directional coverage. \\
\indent Precise characterization of the optical properties of the IceCube detector medium and thereby reducing uncertainties in directional and energy event reconstruction is one of the goals of the IceCube Upgrade. For this purpose, novel calibration devices will be deployed, with the IceCube Upgrade camera system being a key component. 

\subsection{Objective and setup of the camera system}
\begin{figure}
    \centering
    \begin{minipage}{.8\textwidth}
    \centering
    \includegraphics[width=.75\linewidth]{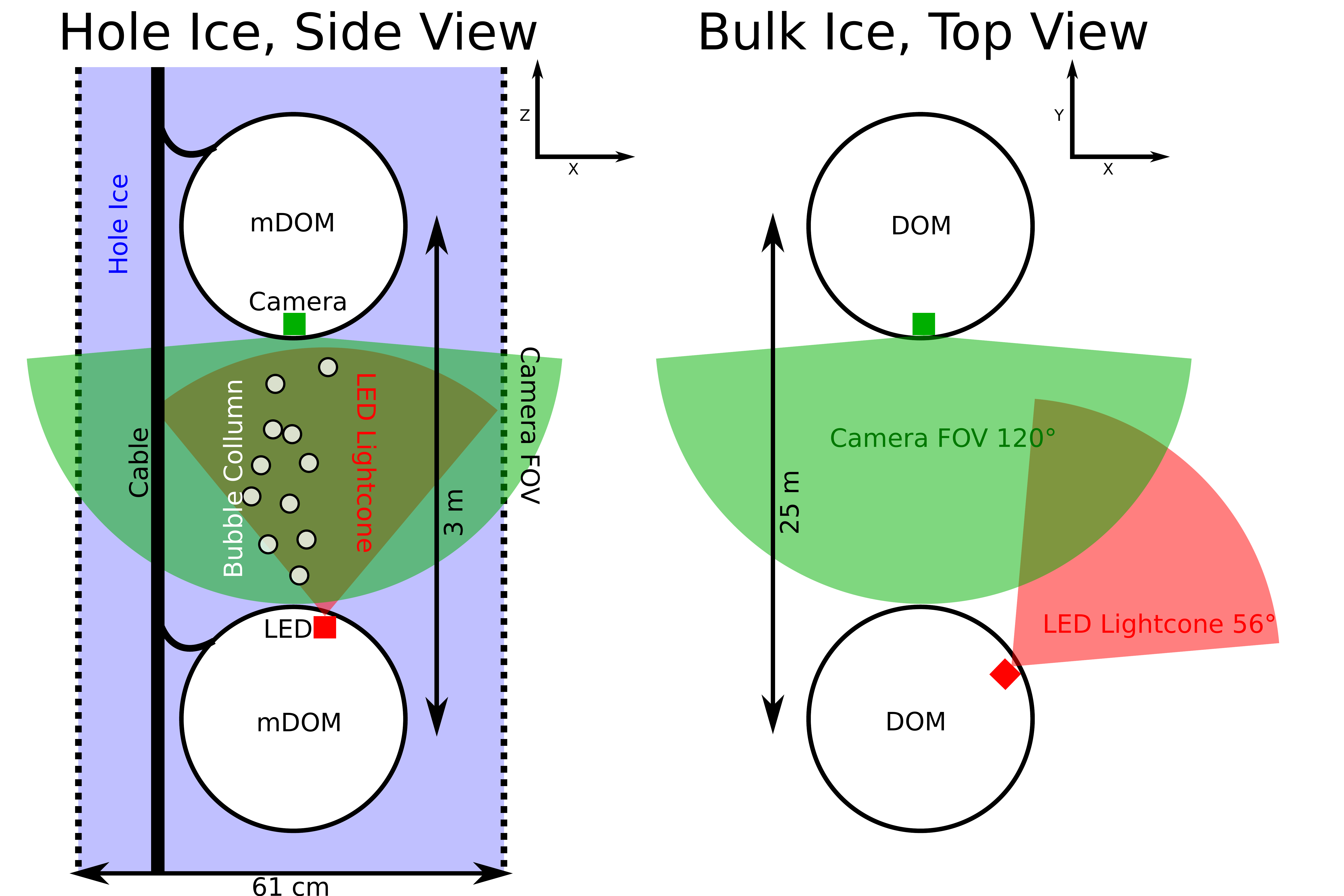}
    \includegraphics[width=.48\linewidth]{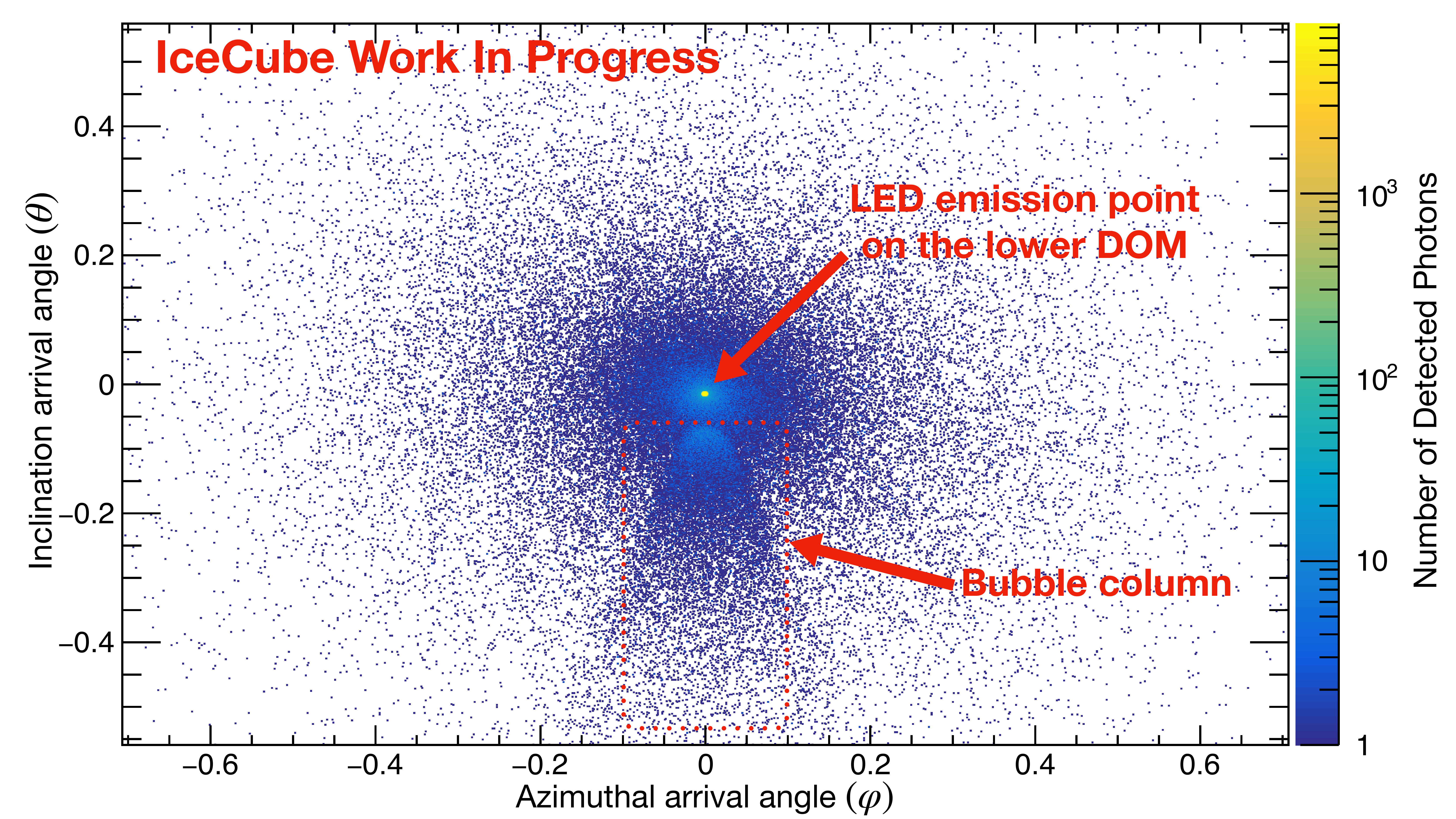}
    \includegraphics[width=.48\linewidth]{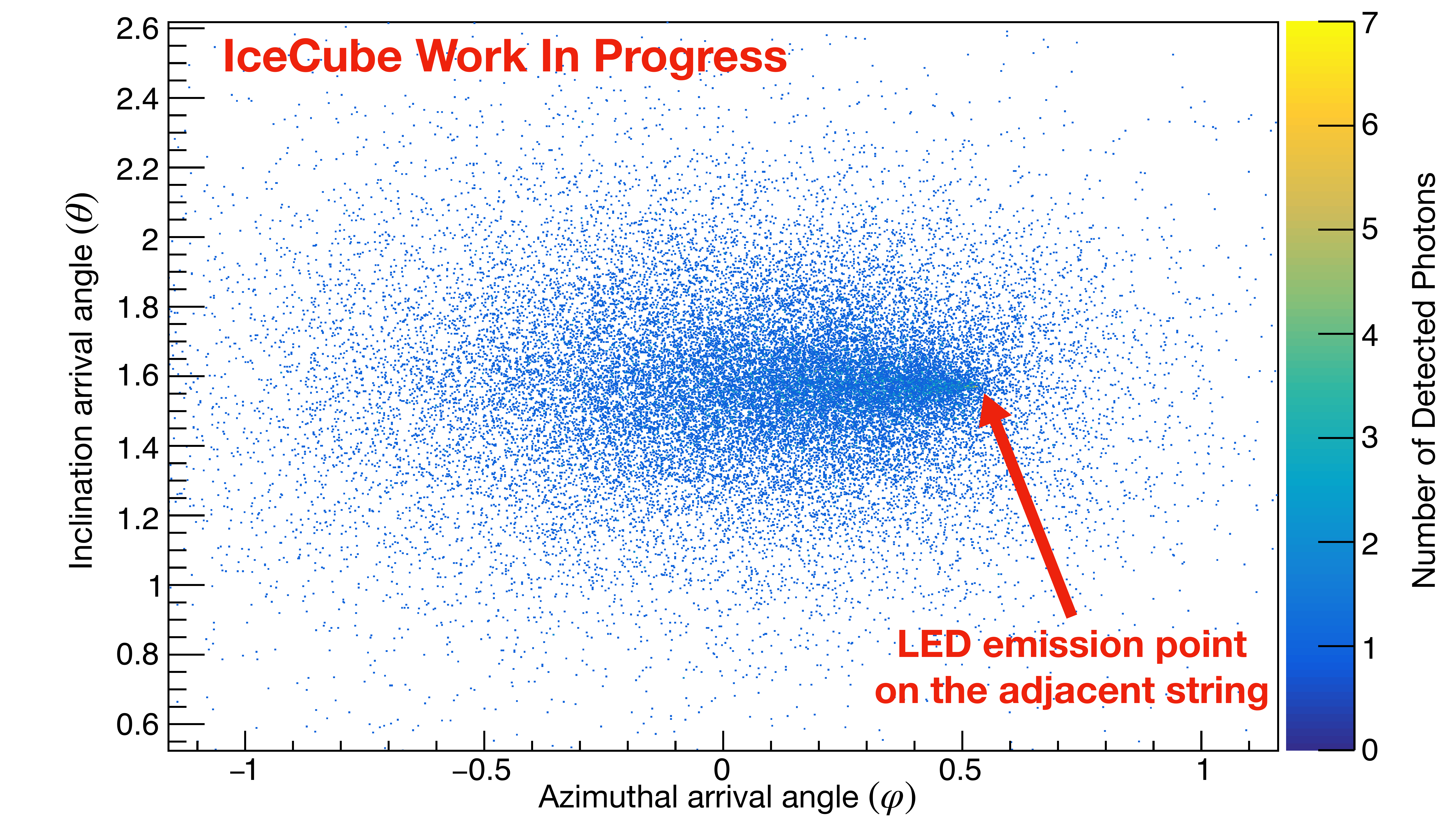}
    \end{minipage}
    \caption[margin=0.8cm]{Schematic of planned IceCube Upgrade camera measurements and the simulated images of them. Left: Refrozen hole ice measurement utilising two vertically separated optical modules on the same string. A downward facing camera observes an upward pointing LED from the DOM below. The simulation image in bottom-left represents imaging the bubble column in the refrozen hole ice. Right: Bulk ice measurement utilising two optical modules on separate strings. A camera is observing scattered light from an LED on an adjacent string pointing at an angle of $60^{\circ}$ to the camera. The plot in the bottom-right shows the light cone observed from an optical module on the adjacent string in the simulation. The schematic is not to scale. The pixel noise is not included in the simulated images.}
    \label{fig:schematic}
\end{figure}
The IceCube Upgrade camera system aims to measure the optical properties of the ice in the vicinity of DOMs. Additionally, information on the position and orientation of the optical sensors will be obtained. To do this, the camera system utilises camera modules integrated inside each newly installed DOM to measure light emitted from illumination modules that accompany each camera pointing in the same direction. With this setup, images of both reflected and transmitted light will be taken.\\
\indent In all three types of new DOMs, three cameras are going to be installed to carry out different types of measurements. In the mDOM two cameras will be installed in the upper hemisphere pointing upwards at $45^{\circ}$. The third camera is positioned at the bottom pole of the mDOM. At the top of the mDOM an additional illumination board is placed to illuminate the refrozen hole ice. In the D-Egg all three cameras are installed on a ring in the lower half of the sensor, point towards the horizon with an angle of $120^{\circ}$ to each other. The method of integration into the pDOMs is currently being developed. \\
\indent In Fig.~\ref{fig:schematic} on the left, one type of measurement is sketched. The downwards facing camera captures direct and scattered light from an illumination module in the DOM below, and the optical properties of the refrozen ice will be inferred based on the distribution of light in the images. Of special interest is a column of ice with different optical properties than the surrounding ice, known as the 'Bubble column', that was originally detected by a special camera system deployed below the deepest DOM of IceCube string~80~\cite{ICdetector}. The other cameras on the mDOMs and D-EGGs measure the optical properties of the bulk ice as shown in Fig.~\ref{fig:schematic}, right. The bulk ice between strings is illuminated by an LED on one of the DOMs, and a camera in a DOM on a neighboring string takes images of the scattered light. The optical properties of the ice can be determined based on the distribution of incident light. Since there are multiple cameras pointing in different directions the bulk ice measurement can be performed direction dependent to gauge the anisotropies in the optical properties of the ice~\cite{ICRC2021:anisotropy}.\\
\indent Simulated images based on a photon propagating Monte Carlo code~\cite{ppc} used in previous camera studies~\cite{ICRC2017:Gen2_camera, ICRC2019:ICU_camera} are shown in Fig.~\ref{fig:schematic}, bottom. The studies are to be extended to develop the image analysis methods which would be applied on the actual image data from the deployed camera system.

\subsection{Hardware}
\begin{figure}
    \centering
    \includegraphics[width=0.85\linewidth]{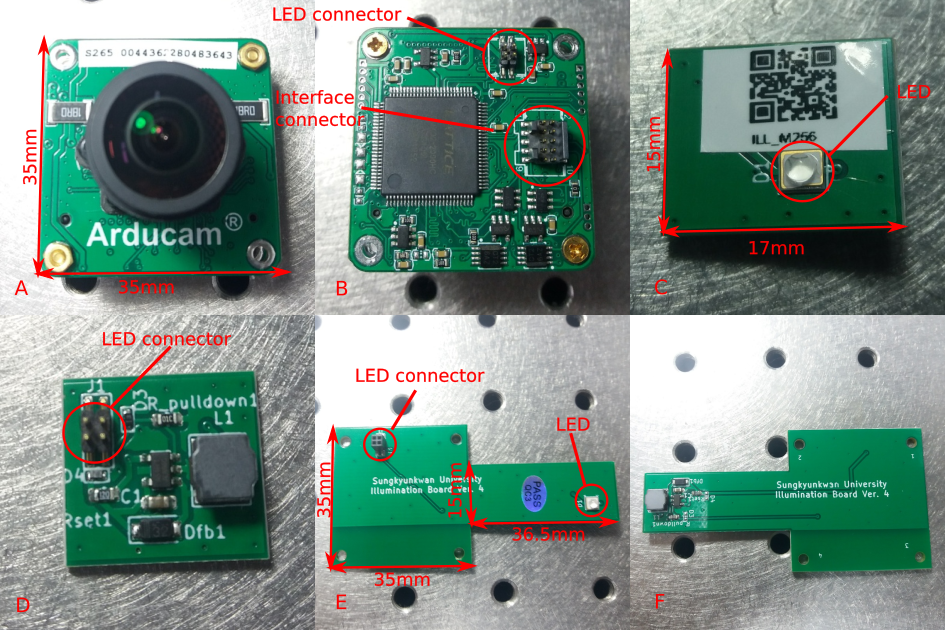}
    \caption[margin=1cm]{The major components of the IceCube Upgrade camera system. A: The camera as seen from the front; B: The camera as seen from the back showing the connectors for the mainboards and the LED system; C: The LED board for the mDOMs from the front with the LED ID code used to identify the boards; D: The LED board backside that shows the connector; E: The D-EGG illumination system from the front showing the board-to board connector and LED; F: The D-EGG illumination system from the back.}
    \label{fig:camdetail}
\end{figure}

\begin{figure}
    \centering
    \begin{minipage}{.80\textwidth}
    \centering
    \includegraphics[width=.40\linewidth]{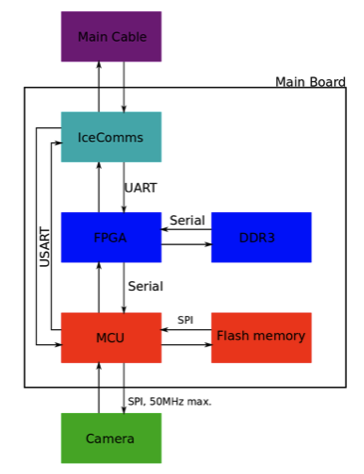}
    \end{minipage}
    \caption[margin=1cm]{Flow chart of camera data and image transfer speeds between camera and DOM main board.}
    \label{fig:comm}
\end{figure}

The camera module for the IceCube Upgrade camera system is a custom designed device consisting of two PCBs constituting the camera and one PCB that serves as an illumination board. The parts can be seen in Fig.~\ref{fig:camdetail}. It uses a Sony \textit{IMX225LQR-C} CMOS image sensor, controlled via a Inter Integrated Circuit~(I2C) interface with a \textit{MachX02} FPGA by Lattice semiconductor. The FPGA also bridges the incoming communication using a Serial Peripheral Interface~(SPI) to the high-speed interface with the image sensor, whose connections to the DOM hardware is shown in Fig.~\ref{fig:comm}. It also extracts the captured image data from the sensor onto an 8~MB RAM inside the camera that serves as a frame buffer. Images have a maximal resolution of 1312 by 979 pixels with a depth of 12 bits resulting in a file size of 2.7~MB per image using full pixel information, which means that the camera can buffer up to 3 images. The illumination board uses an \textit{SSL 80 GB CS8PM1.13} LED from Oslon. The LED is operated with 1~W of power generating 43~lm of light whose dominant wavelength is 470~nm with the spectral bandwidth of 25~nm. The light emitted has a full width at half maximum of $80^{\circ}$.\\
\begin{figure}
    \centering
    \begin{minipage}{0.6\textwidth}
    \centering
    \includegraphics[width=.45\linewidth]{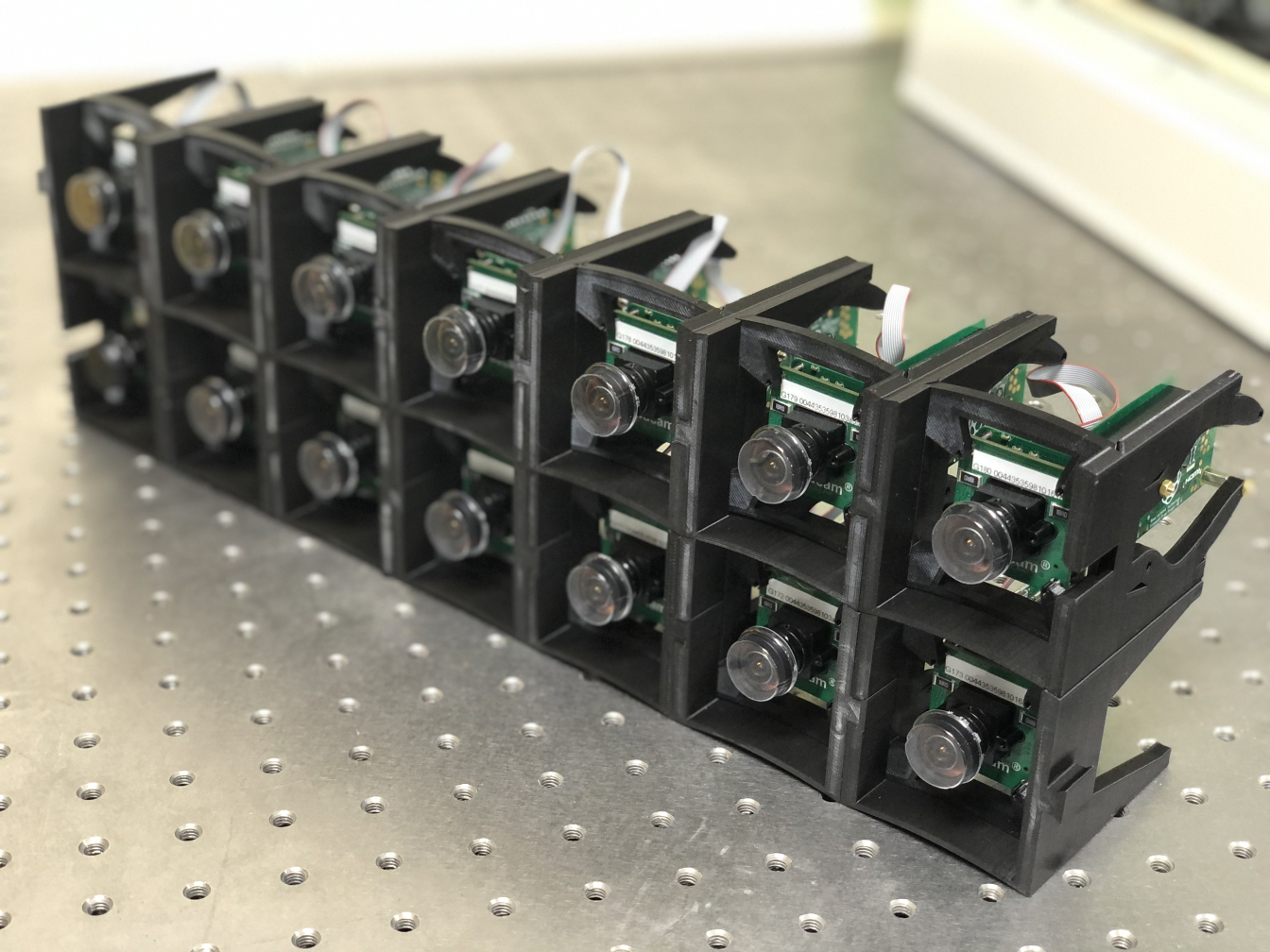}
    \includegraphics[width=.49\linewidth]{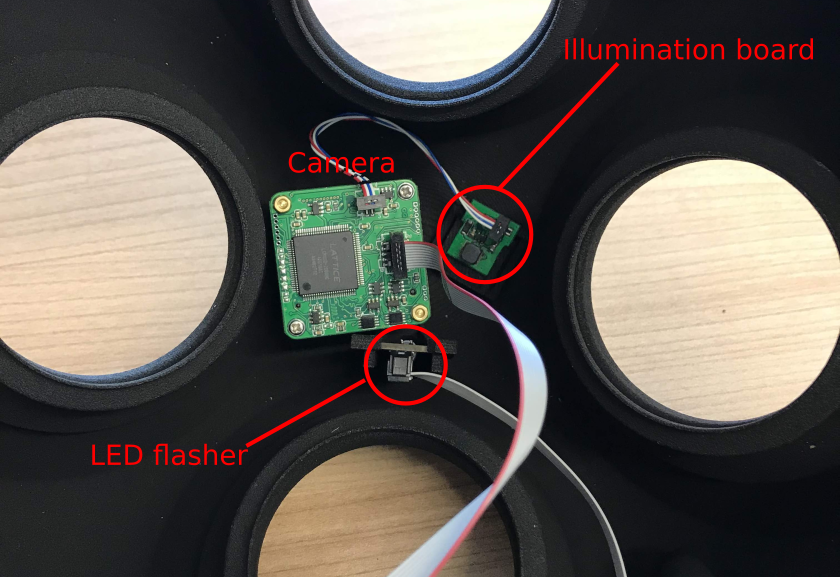}
    \includegraphics[width=.45\linewidth]{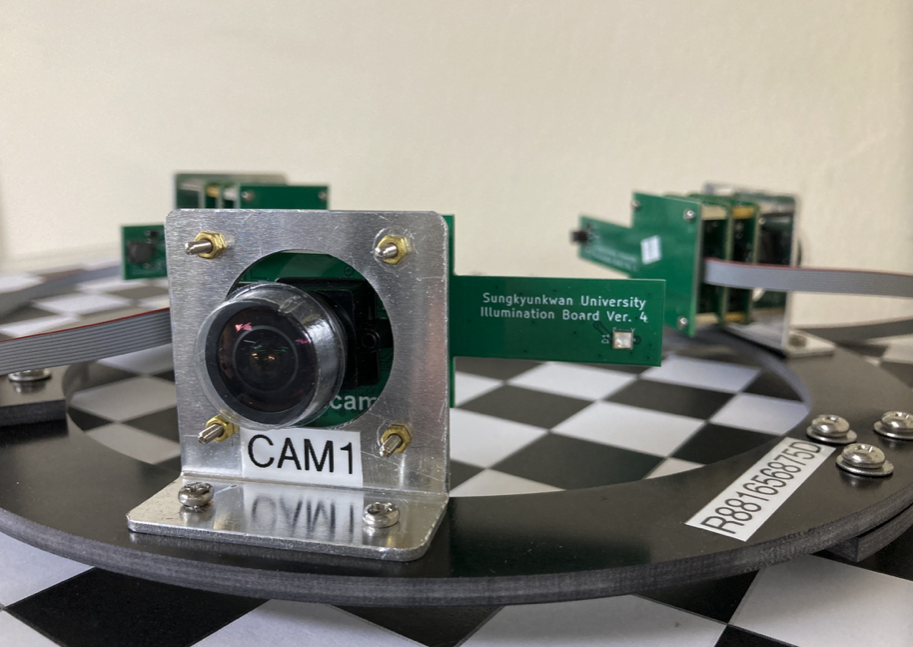}
    \includegraphics[width=.49\linewidth]{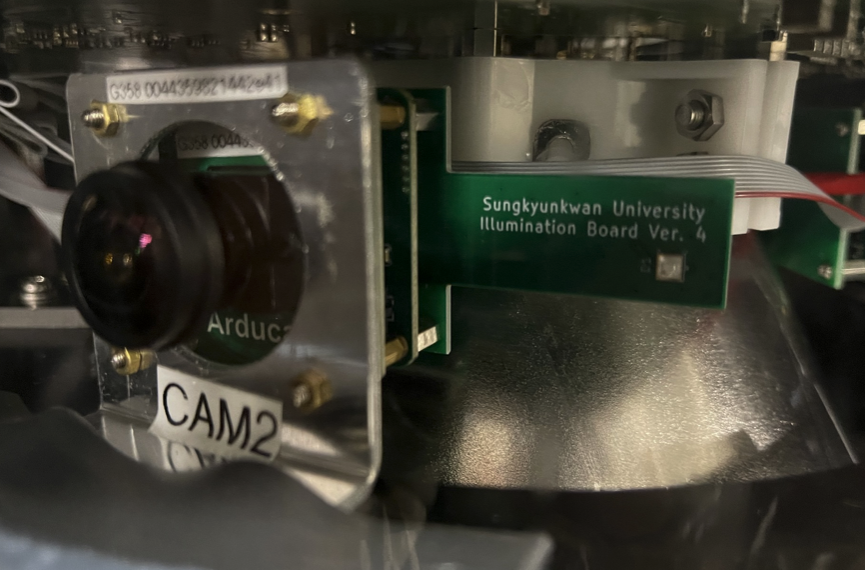}
    \end{minipage}
    \begin{minipage}{0.29\textwidth}
    \includegraphics[width=.94\linewidth]{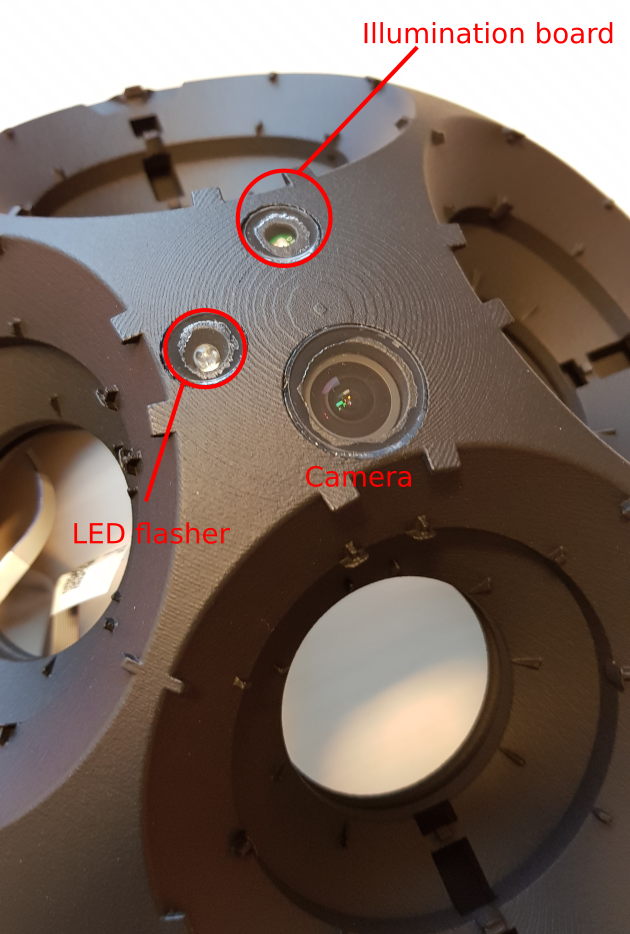}
    \end{minipage} 
    \caption[margin=0.8cm]{Camera testing battery in custom 3D printed holding docks (top left), assembled camera ring for the D-Egg module with three cameras and illumination boards (bottom left), camera ring integrated in a D-EGG (bottom middle), camera inside a mDOM holding structure (top middle), camera in mDOM holding structure seen from the outside (right)}
    \label{fig:camera_photos}
\end{figure}
\indent Cameras for the mDOM are integrated directly into the 3D-printed PMT holding structure. The cameras look through the glass of the pressure vessel using windows in the holding structure. For the D-EGGs, cameras are attached to rings made from fibre reinforced plastic FR-4 using aluminum brackets. An image of such a ring can be seen in Fig.~\ref{fig:camera_photos}. The rings are glued to the glass of the D-EGG pressure vessels using room temperature vulcanizing silicone glue.

\section{Camera calibration and design verification}
\label{sec:CAT}
\begin{figure}
    \centering
    \includegraphics[width=0.8\linewidth]{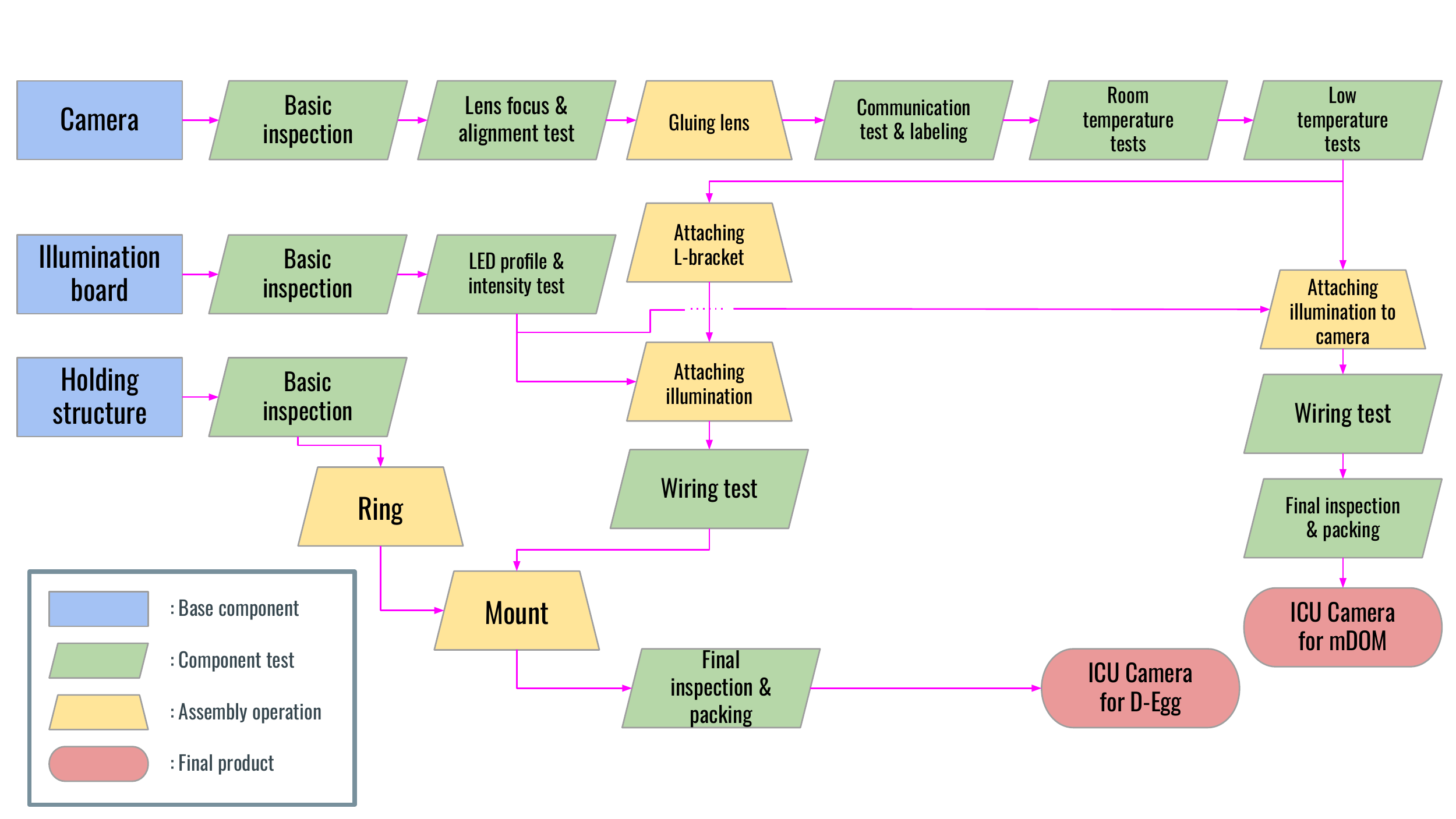}
    \caption[margin=1cm]{Flow chart for the Camera Acceptance Testing.}
    \label{fig:CAT_pipeline}
\end{figure}
To verify the camera operation and to calibrate more than 2000 cameras, all the components are subjected to an extensive suite of tests as shown in Fig.~\ref{fig:CAT_pipeline} (see details in \cite{ICRC2019:ICU_camera}). The entire test cycle for a camera takes 48~hours, with over 3000 (8.1~GB) images captured per camera in the process. During room temperature ($20^{\circ}$C) and low temperature ($-40^{\circ}$C) tests we capture calibration relevant data for each camera.\\
\begin{figure}[ht]
    \centering
    \includegraphics[width=0.75\textwidth]{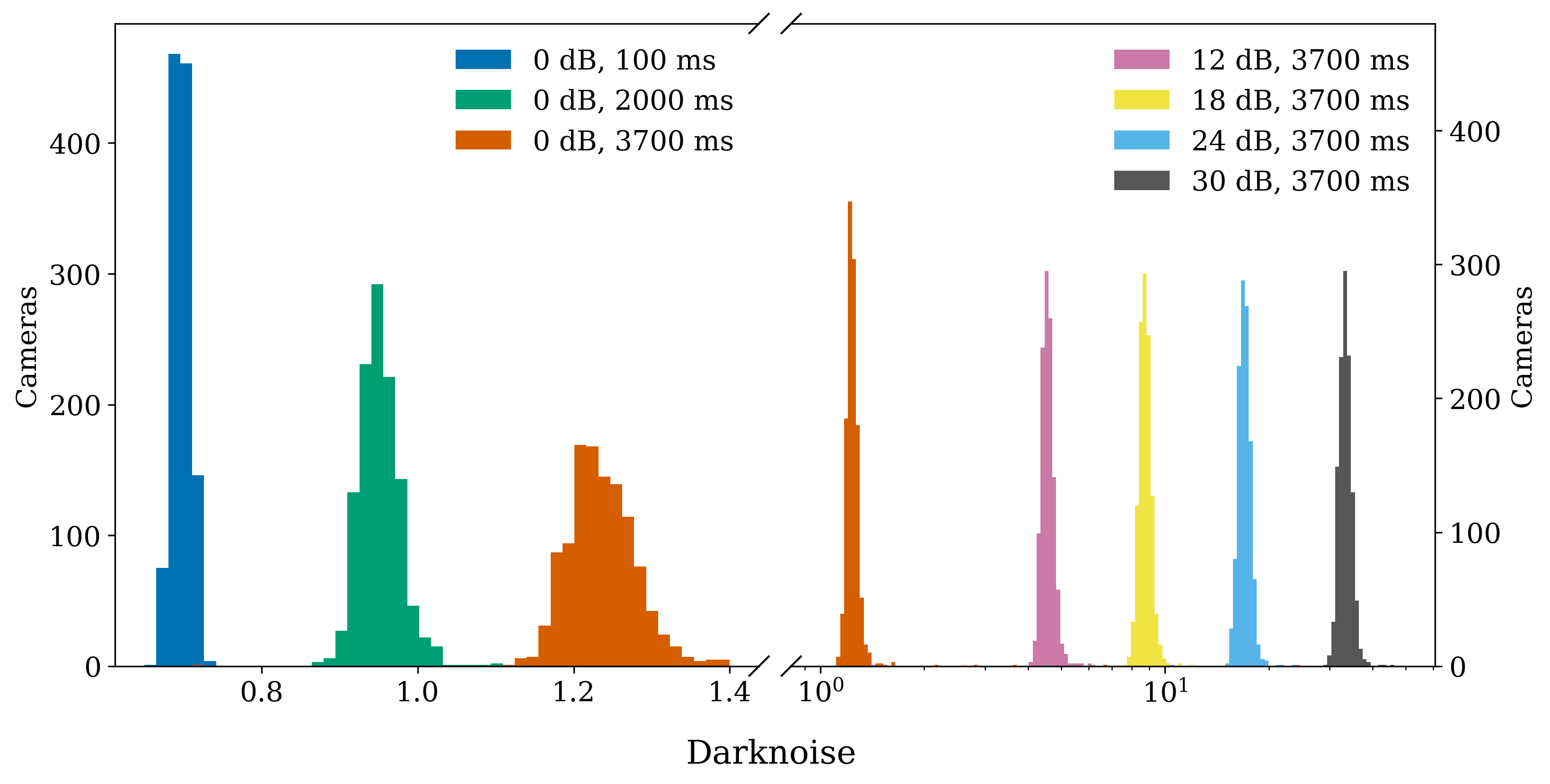}
    \caption{Mean Pixel Darknoise distribution over 1178 cameras for multiple settings}
    \label{fig:darknoise}
\end{figure}
\begin{figure}[ht]
    \centering
    \includegraphics[width=\textwidth]{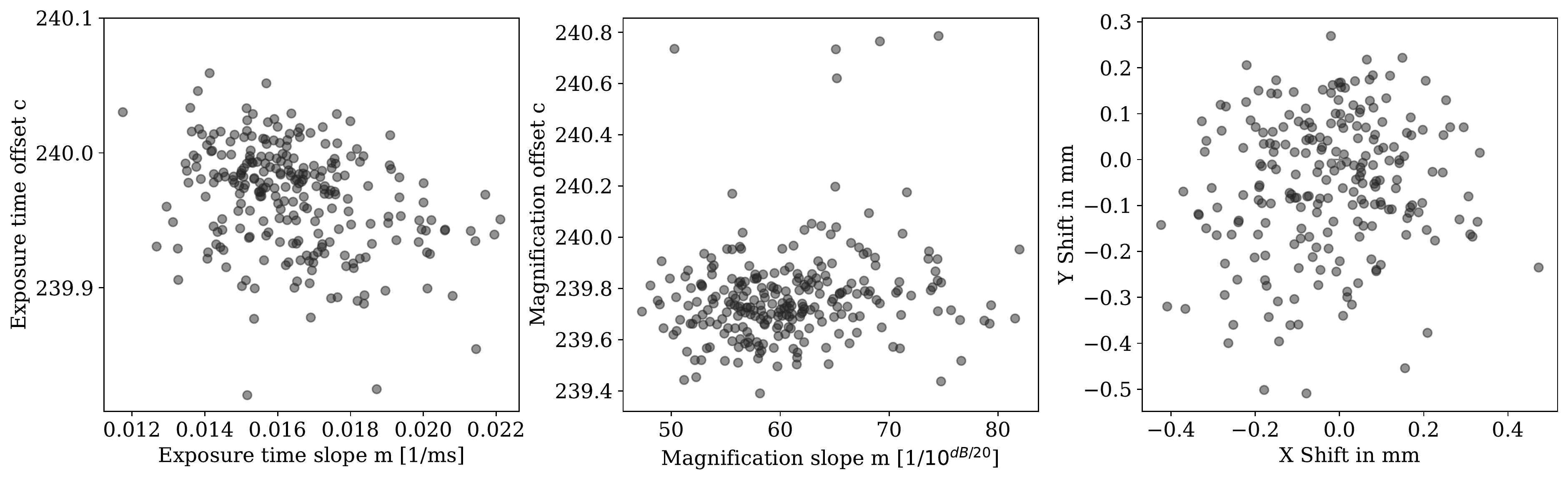}
    \caption{Distribution over 254 cameras, for cameras with R$^2$ of linear fit to exposure time and magnification better than 0.99. Left: Slope m and offset c for a linear fit to pixel response for increasing exposure time. Middle: Same as left, but linear fit for increasing magnification. Right: Shift between image sensor and lens. }
    \label{fig:parameters}
\end{figure}
\indent As cameras will operate in sparse light conditions, a characterization of the pixel darknoise is paramount. Fig.~\ref{fig:darknoise} shows the average pixel darknoise at $-40^{\circ}$C for different camera settings.\\
\indent To verify the camera response to a light source, we take multiple images of an LED at a distance of 1~m at different camera settings. Generally, we find that pixel response for unsaturated pixels scales linearly for each camera with exposure time and magnification, which can be expressed in terms of the camera gain as $\sqrt{10^{Gain\text{[dB]}/10}}$, as shown in Fig.~\ref{fig:parameters}. The camera gain is defined as a factor in the conversion of electric charge per pixel to the digital count of that pixel.\\
\indent We measure the lens - image sensor alignment for each camera (see Fig.~\ref{fig:parameters}, right). Manufacturing inaccuracies can result in a small misalignment between the fish-eye lens and image sensor by $\sim 0.5$~mm. The shift in alignment can be determined with sub pixel accuracy, which translates to an error source for angular estimation below $0.2^{\circ}$.

\section{Run plans for the camera system and application to IceCube-Gen2}
\label{sec:runplan}
\begin{figure}[ht]
    \centering
    \includegraphics[width=0.38\textwidth]{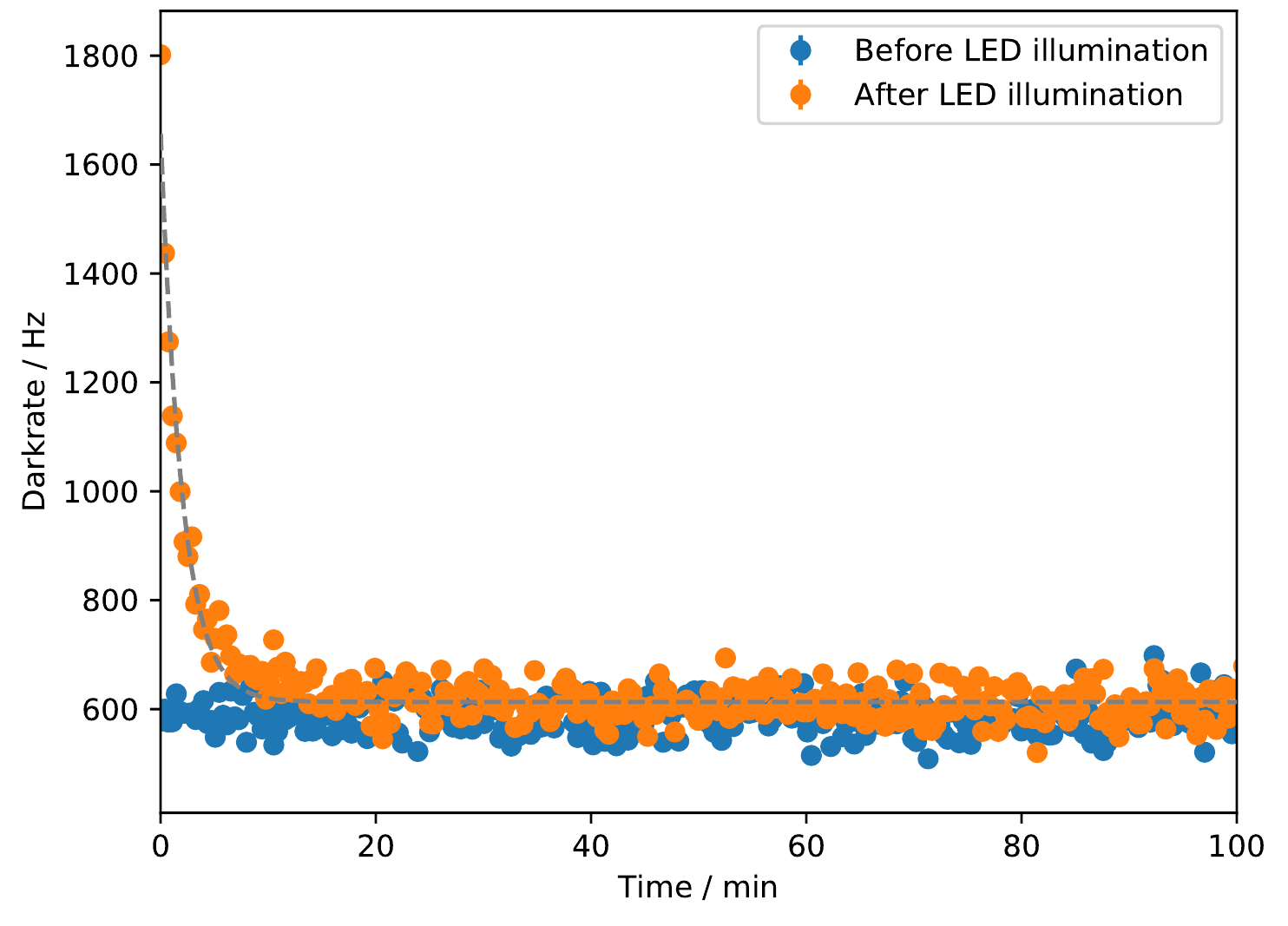}
    \includegraphics[width=0.56\textwidth]{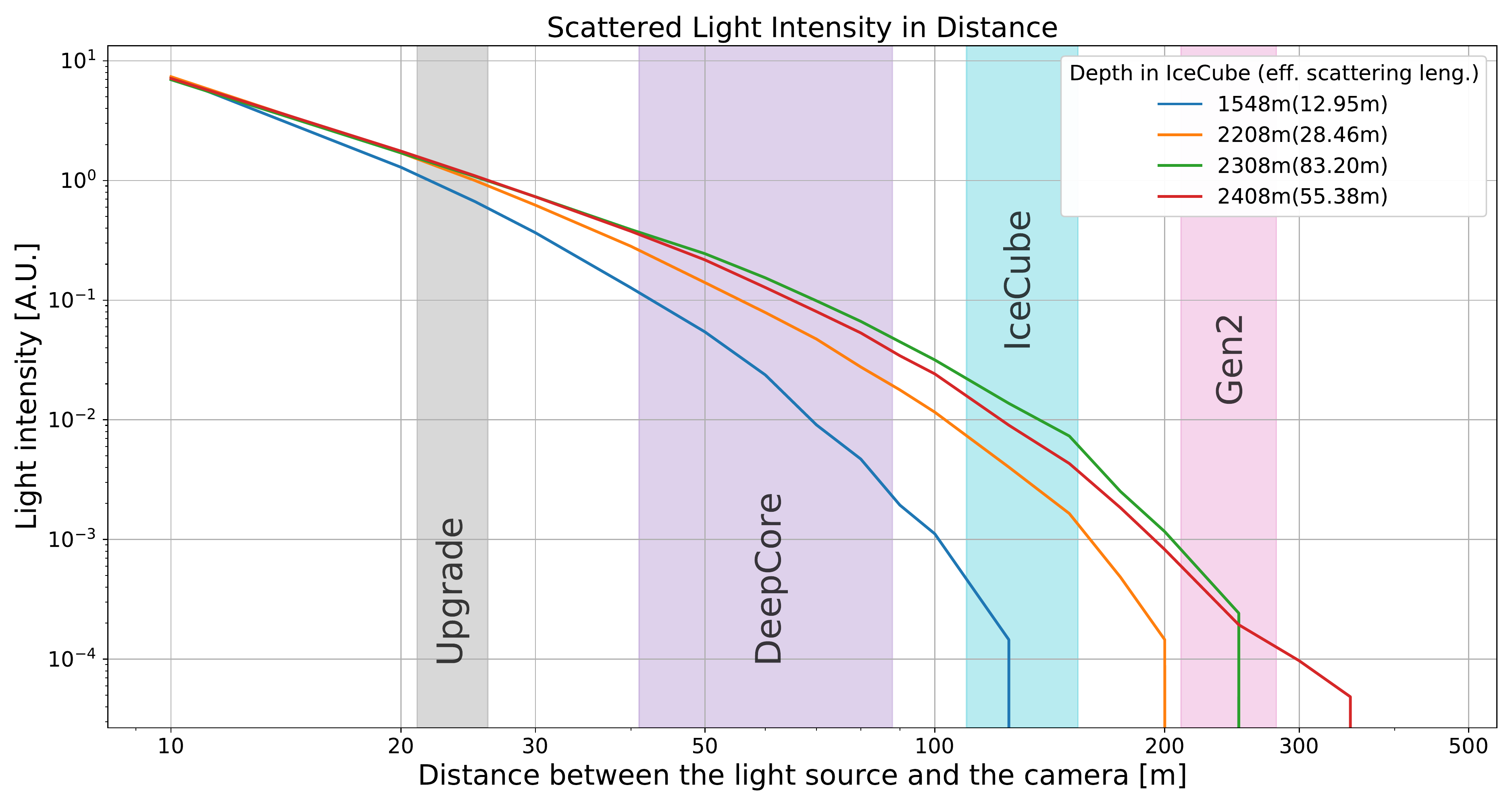}
    \caption{Left: PMT base rate before and after switching on one illumination LED for 90~s. PMT threshold is 0.25~PE. Right: Relative light intensity in cameras directed at an adjacent string with a sideways pointing illumination LED.}
    \label{fig:pmtexcitation}
    \label{fig:photonyield}
\end{figure}
\subsection{In-ice run plan of the camera system}
To maximize science data taking and to minimize the impact on the detector up-time and on the supernova readiness, camera calibration runs will use a number of approaches; these include keeping the minimal run time required to achieve calibration goals and operating of multiple cameras across the detector simultaneously. Camera runs will require the operation of illumination boards that introduce light in the detector (LID), which will require all optical sensor modules to operate in a calibration mode with the PMT HV switched off. Illumination boards will excite PMTs resulting in an increased noise rate following a calibration run. Fig.~\ref{fig:pmtexcitation} shows the increased PMT base rate by switching on the illumination boards and the decrease to normal levels within about 20~min, which is comparable to the settling time of current IceCube detector measured after the runs of the special camera's illumination devices, based on a lab measurement with a 8-inch PMT for D-Egg. Given the low noise rates at the cold operating temperature of IceCube as well as the operational examples from the current IceCube detector, the impact on nominal detector performance is expected to be minimal. The impact on the supernova trigger can be mitigated by excluding optical modules in the vicinity of the illumination boards that were operated during the calibration run.\\
\indent A first set of calibration runs cycling through all cameras and all illumination boards, with one camera or LED operating for every second, will be sufficient to triangulate orientations and positions of all DOMs from the observed LEDs in the set of camera images. In-water tests demonstrated that the camera can resolve 10~cm separations at 25~m distance~\cite{ICRC2019:ICU_camera}.  Once the geometry of each camera is well established, bulk ice measurements can be performed for which a camera observes one or more LEDs on an adjacent string. Hole ice measurements will be performed by operating simultaneously all mDOM - mDOM pairs with the downward facing camera on the upper mDOM taking a transmission photographic image of the illuminated LED in the mDOM below. For reflection photographic hole ice studies a large fraction of all mDOMs will be operated with their bottom camera and the associated LED.

\subsection{Application of camera system to the IceCube-Gen2}
For IceCube-Gen2~\cite{ICRC2021:Gen2} the ice models developed for IceCube can be utilized and a similar camera system base on the experiences for development and operation of the IceCube Upgrade camera system will be employed. The camera system will perform the hole ice related measurements outlined in this work while string to string measurements will be very challenging and are not a priority for Gen2 (see Fig.~\ref{fig:photonyield}). Bulk ice measurements with cameras could be made via back scattered light using setups similar to those utilised for the SPIceCore camera measurements~\cite{ICRC2021:SPIceCore_Camera}.

\section{Conclusions}
\label{conclusions}
The camera system is a key component for a comprehensive understanding of the detector medium. Calibration measurements acquired with the IceCube Upgrade act as a science multiplier as they will enable to retroactively analyze more than 15~years of IceCube data with a substantially improved ice model. Improvements in angular and energy resolution directly affect the science capabilities of IceCube. In particular improved neutrino event pointing is critical for multi-messenger science. A significant fraction of cameras have been tested and integrated into the DOMs for the IceCube Upgrade. The evaluated test data demonstrates the quality of the system and its capabilities. For IceCube-Gen2, a similar camera system will be employed to perform a hole ice survey, which also has a potential for bulk ice studies with back scattered light.

\bibliographystyle{ICRC}
\bibliography{references}

\providecommand{\href}[2]{#2}\begingroup\raggedright\begin{thebibliography}{10}

\bibitem{ICdetector}
{\bfseries IceCube} Collaboration, M.~G. Aartsen {\em et~al.}
  \href{http://dx.doi.org/10.1088/1748-0221/12/03/P03012}{{\em JINST}
  {\bfseries 12} no.~03, (2017) P03012}.

\bibitem{SPICE}
{\bfseries IceCube} Collaboration, M.~G. Aartsen {\em et~al.}
  \href{http://dx.doi.org/10.1016/j.nima.2013.01.054}{{\em Nucl. Instrum. Meth.
  A} {\bfseries 711} (2013) 73--89}.

\bibitem{IceTop}
{\bfseries IceCube} Collaboration, R.~Abbasi {\em et~al.}
  \href{http://dx.doi.org/10.1016/j.nima.2012.10.067}{{\em Nucl. Instrum. Meth.
  A} {\bfseries 700} (2013) 188--220}.

\bibitem{HEnu_Science}
{\bfseries IceCube} Collaboration, M.~G. Aartsen {\em et~al.}
  \href{http://dx.doi.org/10.1126/science.1242856}{{\em Science} {\bfseries
  342} (2013) 1242856}.

\bibitem{TXS}
{\bfseries IceCube} Collaboration, M.~G. Aartsen {\em et~al.}
  \href{http://dx.doi.org/10.1126/science.aat2890}{{\em Science} {\bfseries
  361} no.~6398, (2018) 147--151}.

\bibitem{ICRC2019:ICU-project}
{\bfseries IceCube} Collaboration, A.~Ishihara
  \href{http://dx.doi.org/10.22323/1.358.1031}{{\em PoS} {\bfseries ICRC2019}
  (2021) 1031}.

\bibitem{ICRC2021:D-Egg}
{\bfseries IceCube} Collaboration, C.~Hill {\em PoS} {\bfseries ICRC2021}
  (these proceedings) 1042.

\bibitem{ICRC2021:mDOM}
{\bfseries IceCube} Collaboration, L.~Classen {\em PoS} {\bfseries ICRC2021}
  (these proceedings) 1070.

\bibitem{ICRC2021:anisotropy}
{\bfseries IceCube} Collaboration, M.~Rongen and D.~Chirkin {\em PoS}
  {\bfseries ICRC2021} (these proceedings) 1119.

\bibitem{ppc}
{\bfseries IceCube} Collaboration, D.~Chirkin
  \href{http://dx.doi.org/10.1016/j.nima.2012.11.170}{{\em Nucl. Instrum. Meth.
  A} {\bfseries 725} (2013) 141--143}.

\bibitem{ICRC2017:Gen2_camera}
{\bfseries IceCube-Gen2} Collaboration, M.~Jeong and W.~Kang
  \href{http://dx.doi.org/10.22323/1.301.1040}{{\em PoS} {\bfseries ICRC2017}
  (2018) 1040}.

\bibitem{ICRC2019:ICU_camera}
{\bfseries IceCube} Collaboration, W.~Kang, C.~T\"onnis, and C.~Rott
  \href{http://dx.doi.org/10.22323/1.358.0928}{{\em PoS} {\bfseries ICRC2019}
  (2020) 928}.

\bibitem{ICRC2021:Gen2}
{\bfseries IceCube} Collaboration, A.~Karle and M.~Kowalski {\em PoS}
  {\bfseries ICRC2021} (these proceedings) 22.

\bibitem{ICRC2021:SPIceCore_Camera}
{\bfseries IceCube} Collaboration, D.~Kim {\em PoS} {\bfseries ICRC2021} (these
  proceedings) 1047.

\end{thebibliography}\endgroup

\clearpage
\section*{Full Author List: IceCube Collaboration}

\scriptsize
\noindent
R. Abbasi$^{17}$,
M. Ackermann$^{59}$,
J. Adams$^{18}$,
J. A. Aguilar$^{12}$,
M. Ahlers$^{22}$,
M. Ahrens$^{50}$,
C. Alispach$^{28}$,
A. A. Alves Jr.$^{31}$,
N. M. Amin$^{42}$,
R. An$^{14}$
% ...
\\

\noindent
% ...
$^{12}$ Universit{\'e} Libre de Bruxelles, Science Faculty CP230, B-1050 Brussels, Belgium \\
% ...
$^{14}$ Department of Physics and Laboratory for Particle Physics and Cosmology, Harvard University, Cambridge, MA 02138, USA \\
% ...
$^{17}$ Department of Physics, Loyola University Chicago, Chicago, IL 60660, USA \\
$^{18}$ Dept. of Physics and Astronomy, University of Canterbury, Private Bag 4800, Christchurch, New Zealand \\
% ...
$^{22}$ Niels Bohr Institute, University of Copenhagen, DK-2100 Copenhagen, Denmark \\
% ...
$^{28}$ D{\'e}partement de physique nucl{\'e}aire et corpusculaire, Universit{\'e} de Gen{\`e}ve, CH-1211 Gen{\`e}ve, Switzerland \\
% ...
$^{31}$ Karlsruhe Institute of Technology, Institute for Astroparticle Physics, D-76021 Karlsruhe, Germany  \\
% ...
$^{42}$ Bartol Research Institute and Dept. of Physics and Astronomy, University of Delaware, Newark, DE 19716, USA \\
% ...
$^{50}$ Oskar Klein Centre and Dept. of Physics, Stockholm University, SE-10691 Stockholm, Sweden \\
% ...
$^{59}$ DESY, D-15738 Zeuthen, Germany \\
% ...

\subsection*{Acknowledgements}

\noindent
USA {\textendash} U.S. National Science Foundation-Office of Polar Programs,
U.S. National Science Foundation-Physics Division,
U.S. National Science Foundation-EPSCoR,
Wisconsin Alumni Research Foundation,
Center for High Throughput Computing (CHTC) at the University of Wisconsin{\textendash}Madison,
Open Science Grid (OSG),
Extreme Science and Engineering Discovery Environment (XSEDE),
Frontera computing project at the Texas Advanced Computing Center,
U.S. Department of Energy-National Energy Research Scientific Computing Center,
Particle astrophysics research computing center at the University of Maryland,
Institute for Cyber-Enabled Research at Michigan State University,
and Astroparticle physics computational facility at Marquette University;
Belgium {\textendash} Funds for Scientific Research (FRS-FNRS and FWO),
FWO Odysseus and Big Science programmes,
and Belgian Federal Science Policy Office (Belspo);
Germany {\textendash} Bundesministerium f{\"u}r Bildung und Forschung (BMBF),
Deutsche Forschungsgemeinschaft (DFG),
Helmholtz Alliance for Astroparticle Physics (HAP),
Initiative and Networking Fund of the Helmholtz Association,
Deutsches Elektronen Synchrotron (DESY),
and High Performance Computing cluster of the RWTH Aachen;
Sweden {\textendash} Swedish Research Council,
Swedish Polar Research Secretariat,
Swedish National Infrastructure for Computing (SNIC),
and Knut and Alice Wallenberg Foundation;
Australia {\textendash} Australian Research Council;
Canada {\textendash} Natural Sciences and Engineering Research Council of Canada,
Calcul Qu{\'e}bec, Compute Ontario, Canada Foundation for Innovation, WestGrid, and Compute Canada;
Denmark {\textendash} Villum Fonden and Carlsberg Foundation;
New Zealand {\textendash} Marsden Fund;
Japan {\textendash} Japan Society for Promotion of Science (JSPS)
and Institute for Global Prominent Research (IGPR) of Chiba University;
Korea {\textendash} National Research Foundation of Korea (NRF);
Switzerland {\textendash} Swiss National Science Foundation (SNSF);
United Kingdom {\textendash} Department of Physics, University of Oxford.

\end{document}